\documentclass[aps,prd,reprint,twocolumn,superscriptaddress,showpacs]{revtex4-1}
\usepackage{graphicx}
\usepackage{mathrsfs}
\usepackage{bm}
\usepackage{amsmath}
\usepackage{dcolumn}
\usepackage{epstopdf}
\usepackage{dsfont}
\usepackage{amssymb}
\usepackage{tabularx}
\usepackage{array}
\usepackage{float}
\usepackage{color}
\usepackage{epstopdf}
\usepackage{mathrsfs}
\usepackage[colorlinks, linkcolor=blue,anchorcolor=blue,citecolor=blue,urlcolor=blue]{hyperref}

\begin{document}
\title{Nonsymmorphic-symmetry-protected hourglass Dirac loop, nodal line, and Dirac point in bulk and monolayer $X_3$SiTe$_6$ ($X=$ Ta, Nb)}

\author{Si Li}
\affiliation{Beijing Key Laboratory of Nanophotonics and Ultrafine Optoelectronic Systems, School of Physics,
Beijing Institute of Technology, Beijing 100081, China}
\affiliation{Research Laboratory for Quantum Materials, Singapore University of Technology and Design, Singapore 487372, Singapore}

\author{Ying Liu}\email{ying\_liu@mymail.sutd.edu.sg}
\affiliation{Research Laboratory for Quantum Materials, Singapore University of Technology and Design, Singapore 487372, Singapore}

\author{Shan-Shan Wang}
\affiliation{Research Laboratory for Quantum Materials, Singapore University of Technology and Design, Singapore 487372, Singapore}

\author{Zhi-Ming Yu}
\affiliation{Research Laboratory for Quantum Materials, Singapore University of Technology and Design, Singapore 487372, Singapore}

\author{Shan Guan}
\affiliation{Beijing Key Laboratory of Nanophotonics and Ultrafine Optoelectronic Systems, School of Physics,
Beijing Institute of Technology, Beijing 100081, China}
\affiliation{Research Laboratory for Quantum Materials, Singapore University of Technology and Design, Singapore 487372, Singapore}

\author {Xian-Lei Sheng}
\affiliation{Research Laboratory for Quantum Materials, Singapore University of Technology and Design, Singapore 487372, Singapore}
\affiliation{Department of Physics, Key Laboratory of Micro-nano Measurement-Manipulation and Physics (Ministry of Education), Beihang University, Beijing 100191, China}

\author{Yugui Yao}\email{ygyao@bit.edu.cn}
\affiliation{Beijing Key Laboratory of Nanophotonics and Ultrafine Optoelectronic Systems, School of Physics,
Beijing Institute of Technology, Beijing 100081, China}

\author{Shengyuan A. Yang}
\affiliation{Research Laboratory for Quantum Materials, Singapore University of Technology and Design, Singapore 487372, Singapore}

\begin{abstract}
 Nonsymmorphic space group symmetries can generate exotic band-crossings in topological metals and semimetals. Here, based on symmetry analysis and first-principles calculations, we reveal rich band-crossing features in the existing layered compounds Ta$_3$SiTe$_6$ and Nb$_3$SiTe$_6$, enabled by nonsymmorphic symmetries. We show that in the absence of spin-orbit coupling (SOC), these three-dimensional (3D) bulk materials possess accidental Dirac loops and essential fourfold nodal lines. In the presence of SOC, there emerges an hourglass Dirac loop---a fourfold degenerate nodal loop, on which each point is a neck-point of an hourglass-type dispersion. We show that this interesting type of band-crossing is protected and dictated by the nonsymmorphic space group symmetries, and it gives rise to drumhead-like surface states.  Furthermore, we also investigate these materials in the monolayer form. We show that these two-dimensional (2D) monolayers host  nodal lines in the absence of SOC, and the nodal lines transform to essential spin-orbit Dirac points when SOC is included. Our work suggests a realistic material platform for exploring the fascinating physics associated with nonsymmorphic band-crossings in both 3D and 2D systems.
\end{abstract}

\maketitle
\section{Introduction}
Topological metals and semimetals have been attracting tremendous interest in the current condensed matter physics research~\cite{Chiu2016,Burkov2016,Yang2016,Dai2016,Bansil2016}. In these materials, the electronic band structures exhibit topology/symmetry-protected band-crossings near the Fermi energy, such that the low-energy quasiparticles behave differently from the conventional Schr\"{o}dinger-type fermions, leading to unusual physical properties. As the most prominent examples, Weyl and Dirac semimetals host isolated twofold and fourfold degenerate band-crossing points respectively, around which the electrons resemble the relativistic Weyl and Dirac fermions~\cite{Wan2011,Murakami2007,Burkov2011,Young2012,Wang2012b,Wang2013b,Zhao2013c,Yang2014a,Weng2015,Liu2014c,Borisenko2014,Lv2015,Xu2015a}, giving rise to fascinating effects such as the chiral anomaly~\cite{Nielsen1983,Son2013}. For a three-dimensional (3D) system, besides the 0D nodal points, the nontrivial band-crossings may also take the form of 1D nodal lines~\cite{Fang2016}or even 2D nodal surfaces~\cite{Yang2016,Zhong2016,Liang2016,Bzdusek2017}. The nodal-line materials have been intensively studied recently~\cite{Weng2015c,Yang2014,Mullen2015,Yu2015,Kim2015a,Chen2015,Xie2015,Fang2015,Chen2015a,Chan2016,Li2016,Bian2016,Schoop2016,Gan,Huang2016,Huang2017,Zhang2017,Yu2017,Li2017}. Interesting properties such as the presence of drumhead-like surface states~\cite{Burkov2011a,Yu2017a}, the anisotropic electron transport~\cite{Mullen2015}, the possible surface magnetism/superconductivity~\cite{Chan2016,Wang2017b,Heikkilae2011,Liu2017}, the anomalous Landau level spectrum~\cite{Rhim2015,Lim2017}, and the unusual optical response~\cite{Carbotte2016,Ahn2017,Liu2017a} have been proposed for nodal-line materials.

Depending on their formation mechanism, the various types of band-crossings may be classified into two categories. The first category is for the so-called accidental band-crossings. These crossings are formed by band inversions in certain regions of the Brillouin zone (BZ). They can be removed without changing the symmetry of the system. Examples include the Dirac semimetals Na$_3$Bi and Cd$_3$As$_2$~\cite{Wang2012b,Wang2013b}, the Weyl semimetals in the TaAs family~\cite{Weng2015,Lv2015}, and most proposed nodal-line materials. The second category is for the so-called essential band-crossings. Unlike the accidental ones, the presence of essential band-crossings is guaranteed by the specific space-group symmetry, and they cannot be removed as long as the symmetry is maintained. Nonsymmorphic symmetries, which are operations involving translations with fractional lattice parameters, play a crucial role in generating the essential crossings~\cite{Young2015a,Zhao2016,Yang2017,furusaki2017}, hence such crossings are sometimes also termed as nonsymmorphic band-crossings. The Dirac semimetals BiO$_2$~\cite{Young2012}and some distorted spinels~\cite{Steinberg2014}, the hybrid Dirac metal CaAgBi~\cite{chen2017}, and nodal-line materials ZrSiS~\cite{Schoop2016} are examples hosting essential band-crossings. It was also demonstrated that the nonsymmorphic symmetries may give rise to more exotic types of band-crossings, such as hourglass dispersions~\cite{Wang2016a,Ma2017} and nodal chains~\cite{Bzdusek2016,Wang2017,Yan2017a}.

Currently, it is much desired to search for good candidate materials with essential band-crossings. The motivation is partly due to the fact that the essential band-crossings, if derived from the double representation of the space group, would be intrinsically robust against spin-orbit coupling (SOC). This is especially important for nodal lines, because the accidental band-crossings are typically vulnerable against SOC. So far, good candidate materials with essential band-crossings are still limited. One important point is that although the essential band-crossings are guaranteed to exist in band structure, they may not necessarily appear near the Fermi level. For example, the essential nodal line in ZrSiS is located about 0.5 eV below the Fermi energy~\cite{Schoop2016}. This will severely suppress the manifestation of the band-crossing in electronic properties. The situation is even more challenging for 2D systems, because the structural stability poses more stringent constraints in 2D. Recently, essential 2D Dirac points that are robust against SOC were proposed in monolayer HfGeTe-family materials~\cite{Guan2017}.

In this work, based on symmetry analysis and first-principles calculations, we propose rich topological band-crossings in
the layered ternary telluride compounds Ta$_3$SiTe$_6$ and Nb$_3$SiTe$_6$. The two materials both are existing, and it has been shown that their high-quality single-crystal samples can be synthesized experimentally by the chemical-vapor-transport method~\cite{evain1994,li1992,ohno1999}. We show that in the 3D bulk form, these materials possess an essential fourfold nodal line and an accidental nodal loop in the absence of SOC, and an essential hourglass Dirac loop in the presence of SOC. Interestingly, the hourglass Dirac loop is fourfold degenerate, and is formed by a collection of band-crossing points, each being a neck-point of an hourglass dispersion along some path in a glide mirror plane. We show that the hourglass Dirac loop leads to a pair of spin-split drumhead surface bands. Furthermore, we find that monolayer Ta$_3$SiTe$_6$ and Nb$_3$SiTe$_6$ are also dynamically stable and hence may be realized as 2D materials. These monolayers host essential nodal lines in the absence of SOC; while in the presence of SOC, the nodal line splits and transforms into two essential 2D Dirac points. Importantly, all the band-crossings discussed here are close to the Fermi level. Our result provides a promising material platform for exploring the intriguing properties of essential nodal-line and nodal-point fermions in both 3D and 2D systems.

\section{CRYSTAL STRUCTURE AND First-principles METHODS}

\begin{figure}[b!]
\includegraphics[width=8.2cm]{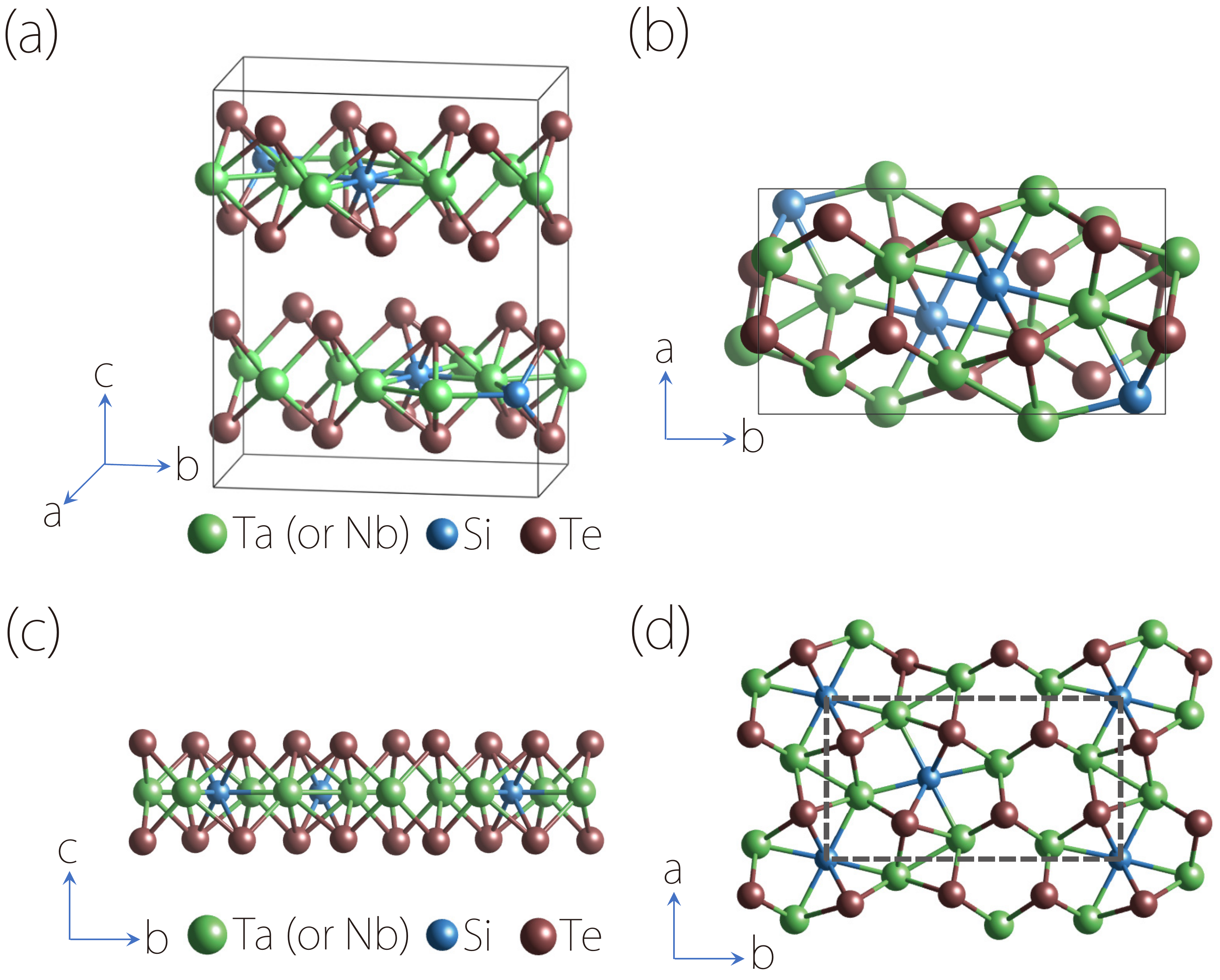}
\caption{(a) Side and (b) top view of the crystal structure of the three-dimensional (3D) bulk Ta$_3$SiTe$_6$ and Nb$_3$SiTe$_6$. The black lines indicate the unit cell. (c) Side and (d) top view of the crystal structure of the monolayer Ta$_3$SiTe$_6$ and Nb$_3$SiTe$_6$. The dashed lines in (d) indicate the unit cell of the monolayer structure.}
\label{fig1}
\end{figure}

Ta$_3$SiTe$_6$ and Nb$_3$SiTe$_6$ are two layered ternary telluride compounds which share an orthorhombic structure with space group No.~62 (Pnma). As shown in Figs.~\ref{fig1}(a) and \ref{fig1}(b), the 3D bulk crystal structure is based upon the stacking of tellurium trigonal prismatic slabs, and each unit cell contains two such slabs that are related by the inversion symmetry. Ta (or Nb) atom are located near the center of a trigonal prism with six Te atoms at the corners~\cite{ohno1999}.
Because of the layered structure, we also study the properties of a single slab (referred to as a monolayer in the following discussion). In fact, few-layer Nb$_3$SiTe$_6$ (thickness $\sim 3$ to $5$ nm) have already been demonstrated in experiment by micro-exfoliation method from the bulk samples~\cite{hu2015}. The structure for a monolayer is shown in Figs.~\ref{fig1}(c) and \ref{fig1}(d). It should be noted that the inversion symmetry that is preserved in the 3D bulk becomes broken for the monolayer structure. This will have important effects on the band structure, as we shall discuss later.

Because the band structure results of Ta$_3$SiTe$_6$ and Nb$_3$SiTe$_6$ share very similar features, in the following discussion, we will focus on the results for Ta$_3$SiTe$_6$. The results for Nb$_3$SiTe$_6$ are presented in the Appendix C.

We have performed first-principles calculations based on the density functional theory (DFT) using the projector augmented wave method as implemented in the Vienna \emph{ab-initio} simulation package~\cite{Kresse1994,Kresse1996,PAW}. The exchange-correlation functional was modeled by the generalized gradient approximation (GGA) with the Perdew-Burke-Ernzerhof (PBE) realization~\cite{PBE}. The van der Waals (vdW) corrections have been taken into account by the approach of Dion \emph{et al.}~\cite{Dion2004}. The cutoff energy was set as 350 eV, and the BZ was sampled with $\Gamma$-centered $k$-mesh of size $8\times 5\times 4$ for the 3D bulk and $9\times 5\times 1$ for monolayer. The structures (for both bulk and monolayer) are fully optimized. The energy and force convergence criteria were set to be $10^{-5}$ eV and $0.01$ eV/\AA, respectively. The optimized lattice parameters for the Ta$_3$SiTe$_6$ 3D bulk structure are $a=6.369$ \AA, $b=11.487$ \AA, and $c=14.109$ \AA, which are close to the experimental values ($a=6.329$ \AA, $b=11.414$ \AA, $c=14.019$ \AA)~\cite{evain1994}. As for the Ta$_3$SiTe$_6$ monolayer structure, the lattice parameters are $a=6.415$ \AA\, and $b=11.568$ \AA\, (a vacuum layer with a thickness of 20 \AA\, was taken to avoid artificial interactions between periodic images). The phonon spectrum is calculated using the PHONOPY code through the DFPT approach~\cite{Togo2015}. As these materials contain transition metal elements (Ta and Nb), the possible correlation effect of the $d$-orbitals was tested via the GGA$+U$ approach~\cite{Anisimov1991,dudarev1998}, which yields almost the same results as those from the GGA calculations (see Appendix B).  The surface states are studied by constructing the maximally localized Wannier functions~\cite{Marzari1997,Souza2001} and by using the iterative Green function method~\cite{Green} as implemented in the WannierTools package~\cite{Wu2017}.

\section{Band Structure for 3D bulk}

The 3D bulk structure of Ta$_3$SiTe$_6$ and Nb$_3$SiTe$_6$ belong to the space group No.~62 (Pnma), which can be generated by the following symmetry elements: the inversion $\mathcal{P}$, the two glide mirrors $\widetilde{\mathcal{M}}_{x}:(x,y,z)\rightarrow (-x+\frac{1}{2},y+\frac{1}{2},z+\frac{1}{2})$, and $\widetilde{\mathcal{M}}_{y}:(x,y,z)\rightarrow (x+\frac{1}{2},-y+\frac{1}{2},z)$. Here the tilde denotes a nonsymmorphic operation, which involves a translation with fractional lattice parameters. By combining the three operations, we also have a third mirror ${\mathcal{M}}_{z}:(x,y,z)\rightarrow (x,y,-z+\frac{1}{2})$, which is a symmorphic operation. In addition, no magnetic ordering has been found for these materials (which we also checked in GGA$+U$ calculations), so the time reversal symmetry $\mathcal{T}$ is also preserved.

In the following Sec.~\ref{3DnoSOC}, we shall first discuss the band structure in the absence of SOC. Then, in Sec.~\ref{3DwSOC}, we shall analyze the result in the presence of SOC.

\subsection{In the absence of SOC: essential fourfold nodal line and accidental nodal loop}\label{3DnoSOC}

\begin{figure}[b!]
\includegraphics[width=8.6cm]{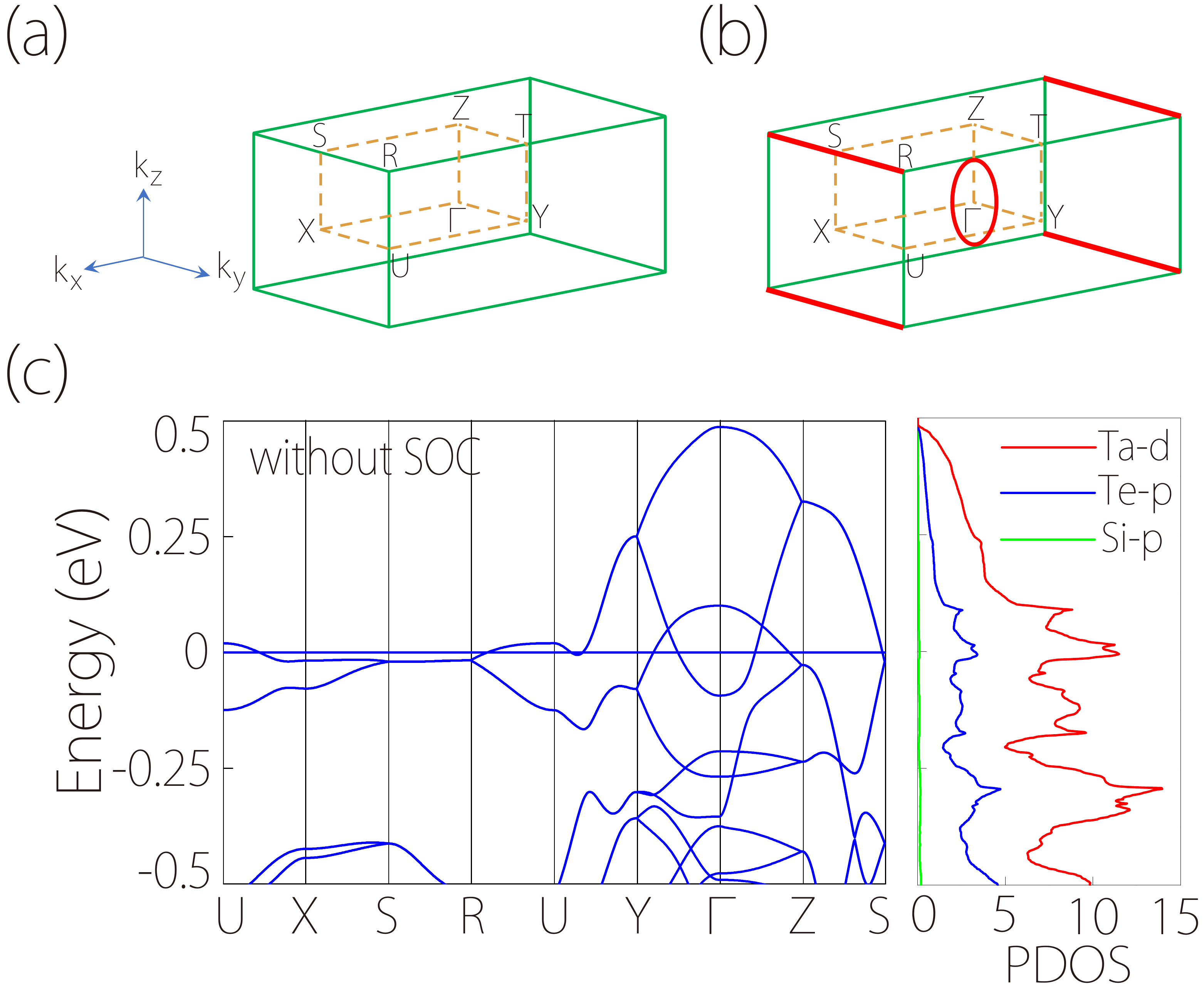}
\caption{(a) Brillouin zone of the 3D bulk. The high-symmetry points are labeled. (b) Schematic figure showing the nontrivial band-crossings in the absence of SOC: the fourfold nodal line on the path S-R (red line) and the accidental nodal loop in the $k_x=0$ plane around the $\Gamma$ point (red circle). (c) Band structure of the three-dimensional Ta$_3$SiTe$_6$ in the absence of SOC. The right panel shows the projected density of states (PDOS).}
\label{fig2}
\end{figure}

We first consider the band structure of 3D Ta$_3$SiTe$_6$ in the absence of SOC, as shown in Fig.~\ref{fig2}(c). From the projected density of states (PDOS), one can observe that the system is metallic and the low-energy states near the Fermi surface are mainly from the Ta $d$-orbitals and Te $p$-orbitals.

There are several nontrivial band features near the Fermi level. First, there are linear band-crossing points along the $\Gamma$-Y and $\Gamma$-Z paths [see Fig.~\ref{fig2}(a) for the BZ of the structure]. In fact, these two points are not isolated. A careful scan shows that they are located on a nodal loop in the $k_x=0$ plane around the $\Gamma$ point, as schematically shown in Fig.~\ref{fig2}(b). Second, the bands along the S-R path form a fourfold degenerate (eightfold degenerate if counting spin) nodal line, and the dispersion along the line is quite flat [see Fig.~\ref{fig2}(c)].

Regarding the nodal-loop in the $k_x=0$ plane, it is formed by the crossing between two bands, and is protected by two independent symmetries: (i) $\widetilde{\mathcal{M}}_x$---the two crossing bands have opposite $\widetilde{\mathcal{M}}_x$ eigenvalues in the $k_x=0$ plane; (ii) $\mathcal{PT}$ symmetry, which requires a quantized $\pi$ Berry phase for a path encircling the loop, hence forbidding the gap opening. These two are typical protection mechanisms for nodal loops in SOC-free systems~\cite{Weng2015c,Mullen2015,Chen2015,Fang2016}. It is interesting that the nodal loop here enjoys a double protection. This nodal loop is accidental, because its formation requires the inverted band ordering between the two crossing bands at $\Gamma$ (compared with that at the BZ boundary) and the loop would be removed if the band inversion does not occur.

Next, we consider the fourfold nodal line along the S-R path with $k_x=\pi$ and $k_z=\pi$ (the wave-vectors are measured in unit of the respective inverse lattice parameters). This nodal line is essential, meaning that its existence is solely dictated by the symmetry, as we analyze in the following.

The S-R path is in the invariant subspace of $\widetilde{\mathcal M}_x$, so each Bloch state $|u\rangle$ on the path there can be chosen as
an eigenstate of $\widetilde{\mathcal M}_x$. One finds that on S-R
\begin{equation}\label{Mx2}
  \widetilde{\mathcal M}_x^2=T_{011}=e^{-ik_y-ik_z},
\end{equation}
where $T_{011}$: $(x,y,z)\rightarrow (x,y+b,z+c)$ is the lattice translation operator. On S-R, $k_z=\pi$, hence the eigenvalues of $\widetilde{\mathcal M}_x$ along this path are $g_x=\pm i e^{-ik_y/2}$.

Note that each $k$ point on S-R is also invariant under the combined anti-unitary operation $\widetilde{\mathcal M}_y\mathcal T$. Since
\begin{equation}
(\widetilde{\mathcal M}_y\mathcal T)^2=e^{-ik_x}=-1
\end{equation}
on S-R, the bands along this path have a Kramer-like double degeneracy. The commutation relation between $\widetilde{\mathcal M}_x$ and $\widetilde{\mathcal M}_y$ is
\begin{equation}\label{MxMy}
  \widetilde{\mathcal M}_x\widetilde{\mathcal M}_y=T_{\bar 1 1 0}\widetilde{\mathcal M}_y\widetilde{\mathcal M}_x.
\end{equation}
On S-R, we have $\widetilde{\mathcal M}_x\widetilde{\mathcal M}_y=-e^{-ik_y}\widetilde{\mathcal M}_y\widetilde{\mathcal M}_x$. Hence, for an eigenstate $|u\rangle$ with $\widetilde{\mathcal M}_x$ eigenvalue $g_x$, we have
\begin{equation}
  \widetilde{\mathcal M}_x(\widetilde{\mathcal M}_y\mathcal T)|u\rangle=g_x(\widetilde{\mathcal M}_y\mathcal T)|u\rangle,
\end{equation}
showing that the two Kramers partners $|u\rangle $ and $(\widetilde{\mathcal M}_y\mathcal T)|u\rangle$ share the same eigenvalue $g_x$.

Additionally, S-R is also the invariant subspace of ${\mathcal M}_z$. The commutation relation between $\widetilde{\mathcal M}_x$ and ${\mathcal M}_z$ is
\begin{equation}
  \widetilde{\mathcal M}_x{\mathcal M}_z=T_{001}{\mathcal M}_z\widetilde{\mathcal M}_x,
\end{equation}
so that $\{\widetilde{\mathcal M}_x,{\mathcal M}_z\}=0$ on S-R, and we have
\begin{equation}
  \widetilde{\mathcal M}_x({\mathcal M}_z|u\rangle)=-g_x({\mathcal M}_z|u\rangle),
\end{equation}
indicating the existence of another degenerate eigenstate $\widetilde{\mathcal M}_z|u\rangle$ which has the opposite eigenvalue $(-g_x)$. Thus, totally, we have four degenerate orthogonal states: $\{|u\rangle, \widetilde{\mathcal M}_y\mathcal T|u\rangle, {\mathcal M}_z|u\rangle, \widetilde{\mathcal M}_y{\mathcal M}_z\mathcal T|u\rangle\}$, forming a degenerate quartet for each $k$ point on S-R. This demonstrates that there is a fourfold (eightfold if spin is counted) degenerate nodal line along S-R when SOC is absent.

It is important to note that the above symmetry analysis is valid when SOC is absent, where $\mathcal{T}^2=+1$ and the rotations (mirrors) operate only on the spatial degrees of freedom. Those symmetry relations are no longer valid if SOC is taken into account, as we discuss in the following section.

\subsection{In the presence of SOC: hourglass Dirac loop}\label{3DwSOC}

\begin{figure}[b!]
\includegraphics[width=8cm]{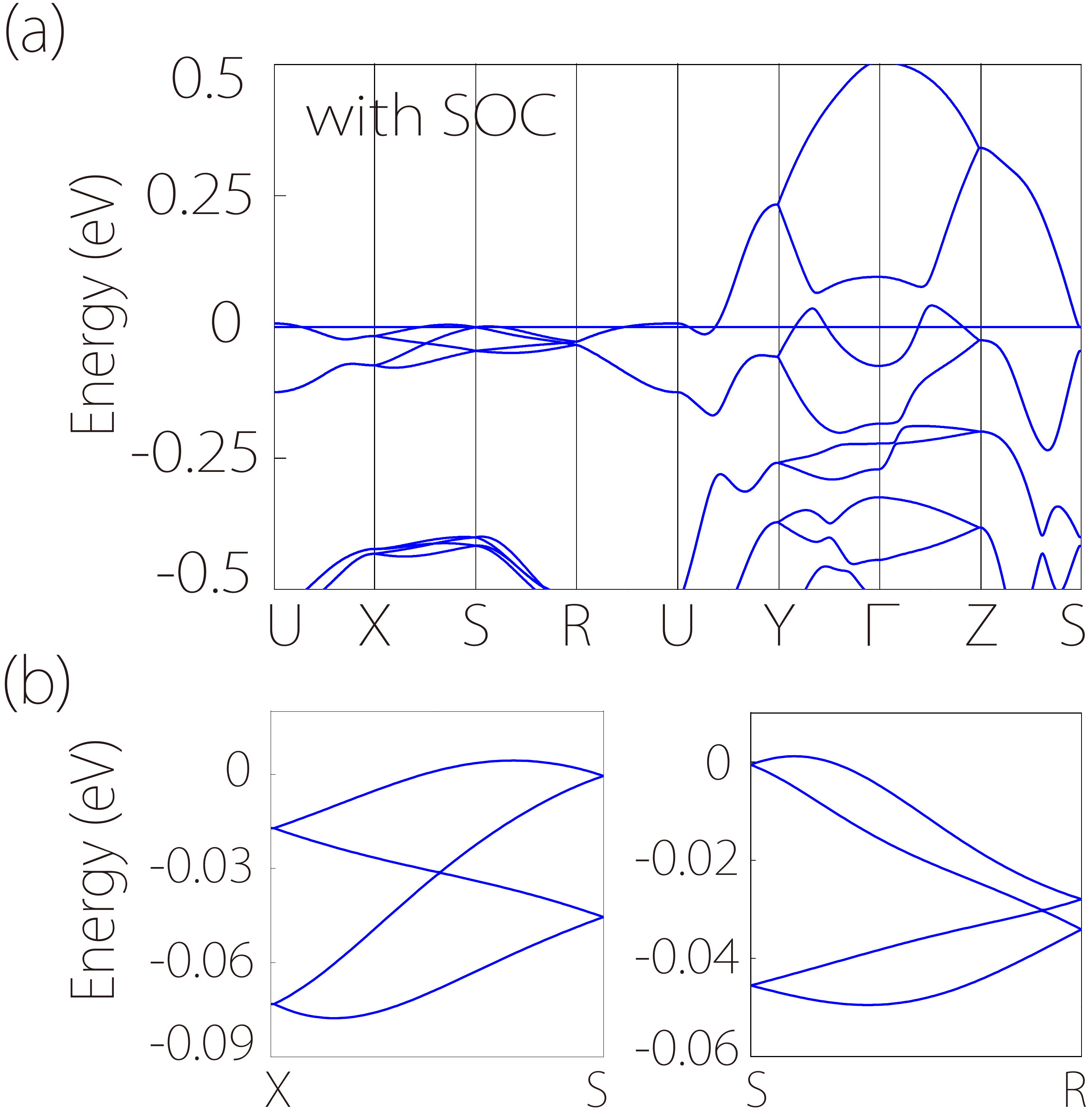}
\caption{(a) Band structure of the 3D Ta$_3$SiTe$_6$ with SOC included. (b) and (c) are the zoom-in images for the low-energy bands along X-S and S-R, showing the hourglass-type dispersions.}
\label{fig3}
\end{figure}

With SOC, the properties of symmetry operations are affected: now we have $\mathcal{T}^2=-1$ and all the rotations (mirrors) need to explicitly operate also on the spin degree of freedom. This strongly modifies the degeneracy and band-crossings in the band structures.

The band structure result of 3D Ta$_3$SiTe$_6$ in the presence of SOC is shown in Fig.~\ref{fig3}(a). Several new features can be observed. First of all, each band is at least twofold degenerate due to the presence of both time reversal and inversion symmetries [with $(\mathcal{PT})^2=-1$]. Second, the nodal loop around the $\Gamma$ point in the $k_x=0$ plane is gapped due to SOC. Third, each band is fourfold degenerate along the paths U-X, R-U, and Z-S.
Last but not least, the nodal line along S-R is also removed, but more interestingly, there emerges an hourglass-type dispersion on S-R [see Fig.~\ref{fig3}(c) for a zoom-in image]. And such hourglass dispersion also appears on the path S-X [Fig.~\ref{fig3}(b)]. Both these two paths are on the $k_x=\pi$ plane.

In the following, we show that the fourfold degeneracy along the paths U-X, R-U, and Z-S, as well as the hourglass dispersion are resulted from the nonsymmorphic symmetries. Let's first consider the fourfold degeneracy on U-X: $(\pi,k_y,0)$, where $-\pi < k_y \le \pi$. This path lies on the $k_x=\pi$ plane which is invariant under $\widetilde{\mathcal{M}}_{x}$, so each Bloch state $|u\rangle$ there can be chosen as an eigenstate of $\widetilde{\mathcal{M}}_{x}$. In the presence of SOC, we have
\begin{equation}\label{eql}
(\widetilde{\mathcal{M}}_{x})^2=T_{011}\overline{E}=-e^{-ik_y}
\end{equation}
on U-X, where $\overline{E}$ denotes the $2\pi$ spin rotation. Compared with Eq.~(\ref{Mx2}) for the spinless case, here we need to explicitly take into account the operation on spin. Hence, the eigenvalue $g_x$ must be $\pm i e^{-ik_y/2}$.

The commutation relation between $\widetilde{\mathcal{M}}_{x}$ and $\mathcal{P}$ on U-X is given by
\begin{equation}\label{MxP}
\widetilde{\mathcal{M}}_{x}\mathcal{P}=T_{111}\mathcal{P}\widetilde{\mathcal{M}}_{x}.
\end{equation}
From Eq.~(\ref{MxP}), one finds that for a state $|u\rangle$ with an $\widetilde{\mathcal{M}}_{x}$ eigenvalue $g_x$, its Kramers partner $\mathcal{PT}|u\rangle$ satisfies
\begin{equation}\label{eq3}
\widetilde{\mathcal{M}}_{x}(\mathcal{P}\mathcal{T}|u\rangle)
=g_x(\mathcal{P}\mathcal{T}|u\rangle).
\end{equation}
This shows that $|u\rangle$ and $\mathcal{P}\mathcal{T}|u\rangle$ have the same $g_x$.

On the other hand, path U-X is also the invariant space of the anti-unitary symmetry $\widetilde{\mathcal M}_y\mathcal T$, and one can show that $(\widetilde{\mathcal M}_y\mathcal T)^2=-1$ on U-X, also leading to a Kramers-like degeneracy, i.e., each state $|u\rangle$ also has a degenerate partner $\widetilde{\mathcal M}_y\mathcal T|u\rangle$. In the presence of SOC, the commutation relation in Eq.~(\ref{MxMy}) gets modified to
\begin{equation}
  \widetilde {\mathcal M}_x\widetilde{\mathcal M}_y=-T_{\bar 110}\widetilde {\mathcal M}_y\widetilde{\mathcal M}_x,
\end{equation}
where the negative sign is due to the operation on spin, i.e., from $\left\{\sigma_x,\sigma_y\right\}=0$. Hence, on the U-X path, one finds that $\widetilde {\mathcal M}_x\widetilde{\mathcal M}_y=e^{-ik_y}\widetilde {\mathcal M}_y\widetilde{\mathcal M}_x$, and
\begin{equation}
  \widetilde {\mathcal M}_x(\widetilde{\mathcal M}_y\mathcal T|u\rangle)=-g_x(\widetilde{\mathcal M}_y\mathcal T|u\rangle).
\end{equation}
This demonstrates that the two states $\widetilde{\mathcal M}_y\mathcal T|u\rangle$ and $|u\rangle$ have opposite eigenvalues $g_x$. Therefore, for each $k$ point on U-X, there must be a fourfold degenerate quartet $\{|u\rangle,\mathcal{PT}|u\rangle, \widetilde{\mathcal M}_y\mathcal T|u\rangle, \mathcal P\widetilde{\mathcal M}_y|u\rangle\}$. The degeneracies on the R-U and Z-S paths can be argued in a similar way. The detailed analysis is presented in the Appendix A.

Next, we turn to the hourglass-type dispersion on the path S-X: $(\pi,0,k_z)$. Since this path also lies on the $k_x=\pi$ plane, each Bloch state $|u\rangle$ there can also be chosen to have an well-defined eigenvalue $g_x=\pm i e^{-ik_y/2-ik_z/2}$. Following similar analysis above, one finds that on S-X,
\begin{equation}
\widetilde{\mathcal{M}}_{x}(\mathcal{P}\mathcal{T}|g_x\rangle)=g_x(\mathcal{P}\mathcal{T}|g_x\rangle).
\end{equation}
Hence, the Kramers partners $|u \rangle$ and $\mathcal{P}\mathcal{T}|u \rangle$ on S-X share the same eigenvalue $g_x$.

\begin{figure}[t!]
\includegraphics[width=8.6cm]{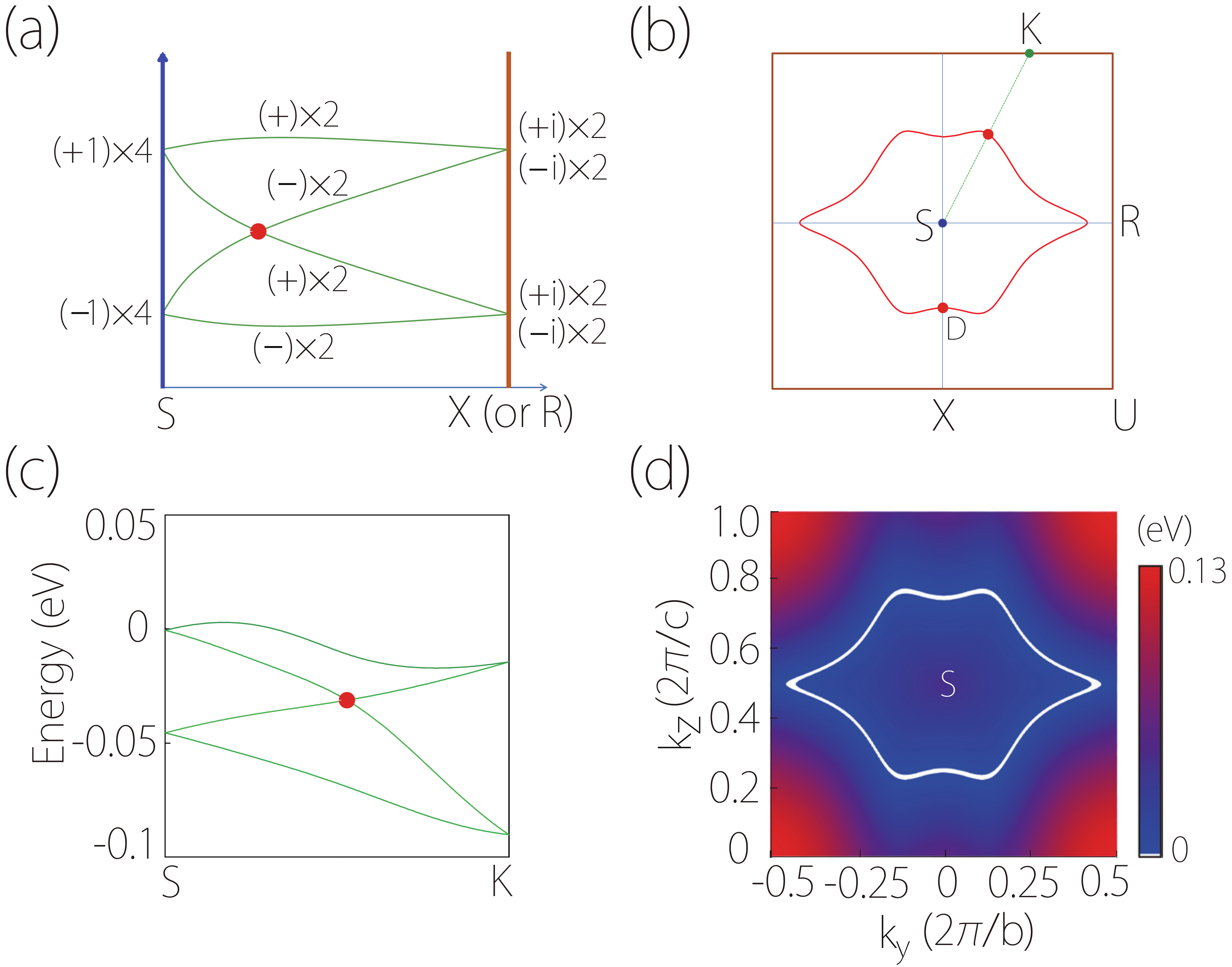}
\caption{(a) Schematic figure showing the hourglass dispersion along an arbitrary path on the $k_x=\pi$ plane connecting S to the point X (or R). The numbers in the brackets indicate the eigenvalues $g_x$ of the states. (b) Schematic figure showing that the neck-point of the hourglass dispersion [the red point in (a)] traces out a fourfold degenerate Dirac loop on the $k_x=\pi$ plane. (c) DFT result for the hourglass dispersion on path S-K, where K is the mid-point between X and U. (d) Shape of the hourglass Dirac loop (the white colored loop) obtained from the DFT calculations. The color-map indicates the local gap between the two crossing bands.}
\label{fig4}
\end{figure}

Meanwhile, S and X are at time reversal invariant momenta (TRIM), which are invariant under $\mathcal{T}$.
At S: $(\pi,0,\pi)$, since $g_x=\pm 1$, each Kramers pair $|u\rangle$ and $\mathcal{T}|u\rangle$ at S must have the same eigenvalues $g_x$. And so the degenerate quartet at S (which may be chosen as $\left\{|u\rangle,\mathcal{T}|u\rangle,\mathcal{P}|u\rangle,\mathcal{P}\mathcal{T}|u\rangle\right\}$) must have the same $g_x$. On the other hand, at point X: $(\pi,0,0)$, since $g_x=\pm i$, each Kramers pair $|u\rangle$ and $\mathcal{T}|u\rangle$ must have opposite $g_x$, and the degenerate quartet must consist of two states with $g_x = +i$ and two other states with $g_x = -i$. Thus, there must be a partner-switching when going from S to X, which leads to the hourglass-type dispersion, as schematically illustrated in Fig.~\ref{fig4}(a). Importantly, the neck-point of the hourglass is fourfold degenerate, and it is protected because the two crossing bands (each has a double degeneracy due to $\mathcal{PT}$) have the opposite $g_x$. The similar analysis also applies to the path S-R.

Furthermore, the above argument can be generalized to apply for an arbitrary path on the $k_x=\pi$ plane connecting S to a point on the boundary lines U-X and R-U [see Fig.~\ref{fig4}(b)]. For example, Fig.~\ref{fig4}(c) shows the hourglass spectrum obtained for a path connecting S to a point K on U-X. Essentially, this is because the $k_x=\pi$ plane is invariant under $ \widetilde {\mathcal M}_x$, and the states on the boundary line U-X are fourfold degenerate with $g_x$ eigenvalues paired as two with $g_x=+ie^{-ik_y/2}$ and the other two with $g_x=-ie^{-ik_y/2}$. Thus, a partner-switching and hence the hourglass dispersion are guaranteed when going from S to a point on U-X (and also R-U by a similar analysis).

Remarkably, as a consequence of the argument, the neck-point of the hourglass dispersion must trace out a closed Dirac loop surrounding point S on the $k_x=\pi$ plane [see Fig.~\ref{fig4}(b)]. This is indeed the case, as confirmed by our DFT result as shown in Fig.~\ref{fig4}(d). Our analysis demonstrates that this hourglass Dirac loop is essential and solely dictated by the nonsymmorphic space group symmetry.

\begin{figure}[t!]
\includegraphics[width=8.6cm]{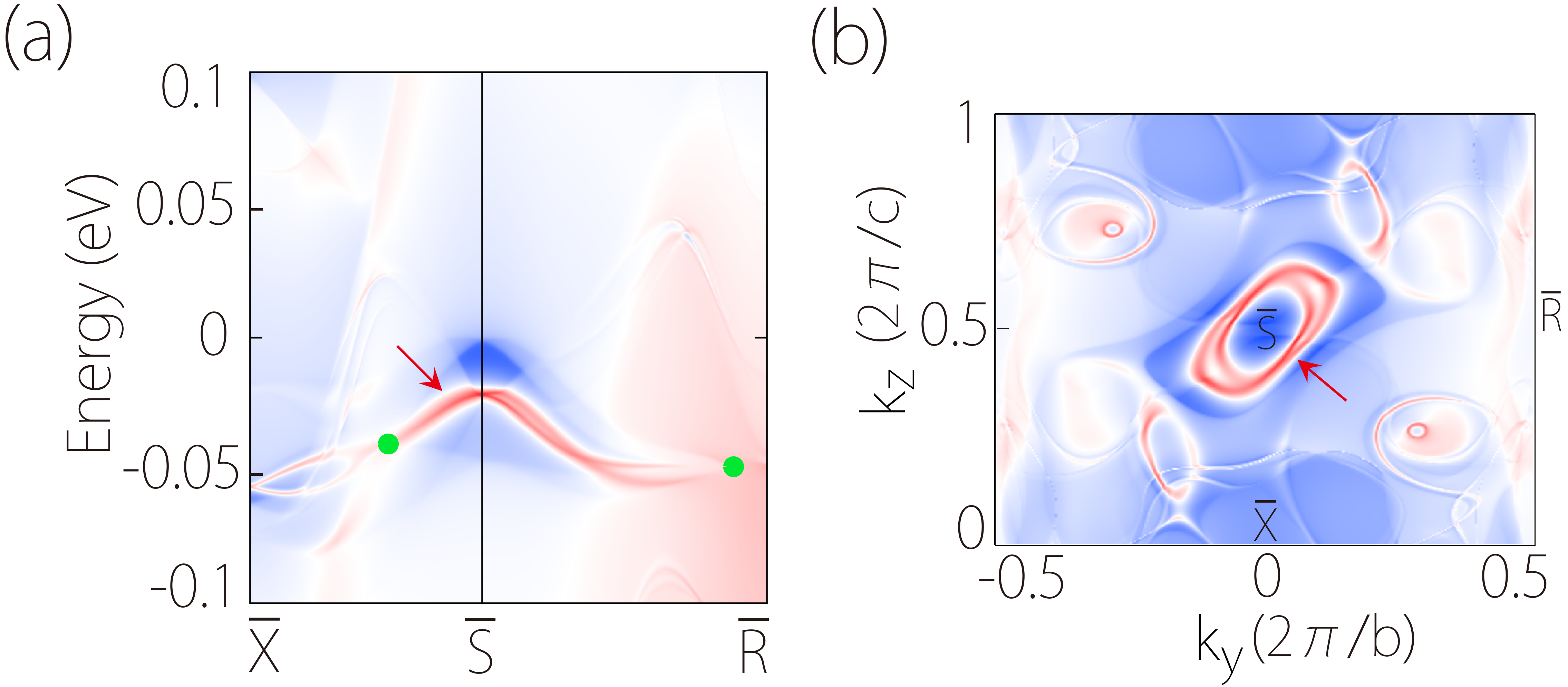}
\caption{(a) Projected spectrum on the (100) surface of 3D Ta$_3$SiTe$_6$. (b) Constant energy slice at $-0.03$ eV. The green dots in
(a) mark the location of the projected bulk band-crossing points on the hourglass Dirac loop. The arrows in (a) and (b) indicate the drumhead-like surface states.}
\label{fig5}
\end{figure}

It has been shown that the nodal loop in the bulk often lead to
drumhead-like surface states at the sample surface where the loop has a nonzero projected area~\cite{Weng2015c,Yang2014}. In Fig.~\ref{fig5}, we plot the surface spectrum of the (100) surface. One observes that there indeed appear a pair of drumhead-like surface bands within the projected bulk loop. The drumhead-like surface bands are split, because the inversion symmetry is broken at the surface such that the $\mathcal{PT}$-enforced double degeneracy is lifted for the surface bands. This feature is similar to that found in the hourglass Dirac chain material ReO$_2$~\cite{Wang2017}.

\section{Results for 2D monolayer}

Since Ta$_3$SiTe$_6$ and Nb$_3$SiTe$_6$ have a layered structure, it is natural to consider their monolayer or few-layer structures. Here we focus on the monolayer structure. We expect this case to be most interesting because it loses the inversion symmetry, which may lead to features contrasting with the bulk band structure.

The fully optimized lattice structure for the monolayer is shown in Figs.~\ref{fig1}(c) and \ref{fig1}(d). To verify its dynamical stability, we performed phonon spectrum calculation. The obtained phonon spectrum is plotted in Fig.~\ref{fig6}(b).
One observes that there is no pronounced imaginary frequency in the spectrum, showing the dynamical stability of the structure. Around the $\Gamma$ point, besides the two linearly dispersing in-plane transverse acoustic branches, there is also the parabolic out-of-plane acoustic (ZA) branch, which is a characteristic feature of 2D materials~\cite{Liu2007,Zhu2014}. We mention that the ZA branch goes slightly below zero near $\Gamma$ along the $\Gamma$-Q path. This is due to the computational error originated from the finiteness in the supercell size and the $k$-mesh size used in the numerical calculation. Physically, since the acoustic modes near $\Gamma$ correspond to the long wavelength limit, its accurate description necessarily requires a large supercell size and a dense $k$-mesh in the calculation. For the monolayer structure studied here, the number of atoms in the unit cell is quite large (20 atoms/unit cell), the maximum size we can afford to compute is for a $3\times 2\times 1$ supercell with a $3\times 3\times 1$ $k$-mesh, which cannot completely eliminate the appearance of negative frequencies near $\Gamma$. However, we have checked that with increasing size of the supercell and/or $k$-mesh, the negative-frequency part is gradually removed. Hence, we expect that the slightly negative frequencies near $\Gamma$ in Fig.~\ref{fig6}(b) is not physical and should be eliminated when the supercell as well as the $k$-mesh sizes are sufficiently large.

Without inversion symmetry, the space group for the monolayer structure becomes No.~26 (Pmc21), which may be generated by the two elements: a glide mirror $\widetilde{\mathcal{M}}_{y}:(x,y,z)\rightarrow(x+\frac{1}{2},-y,z)$ and a mirror ${\mathcal{M}}_{z}:(x,y,z)\rightarrow(x,y,-z)$. In the following, we shall first analyze the band structure in the absence of SOC, and then discuss the result with SOC included.

\subsection{In the absence of SOC: essential nodal line}

\begin{figure}[t!]
\includegraphics[width=8.8cm]{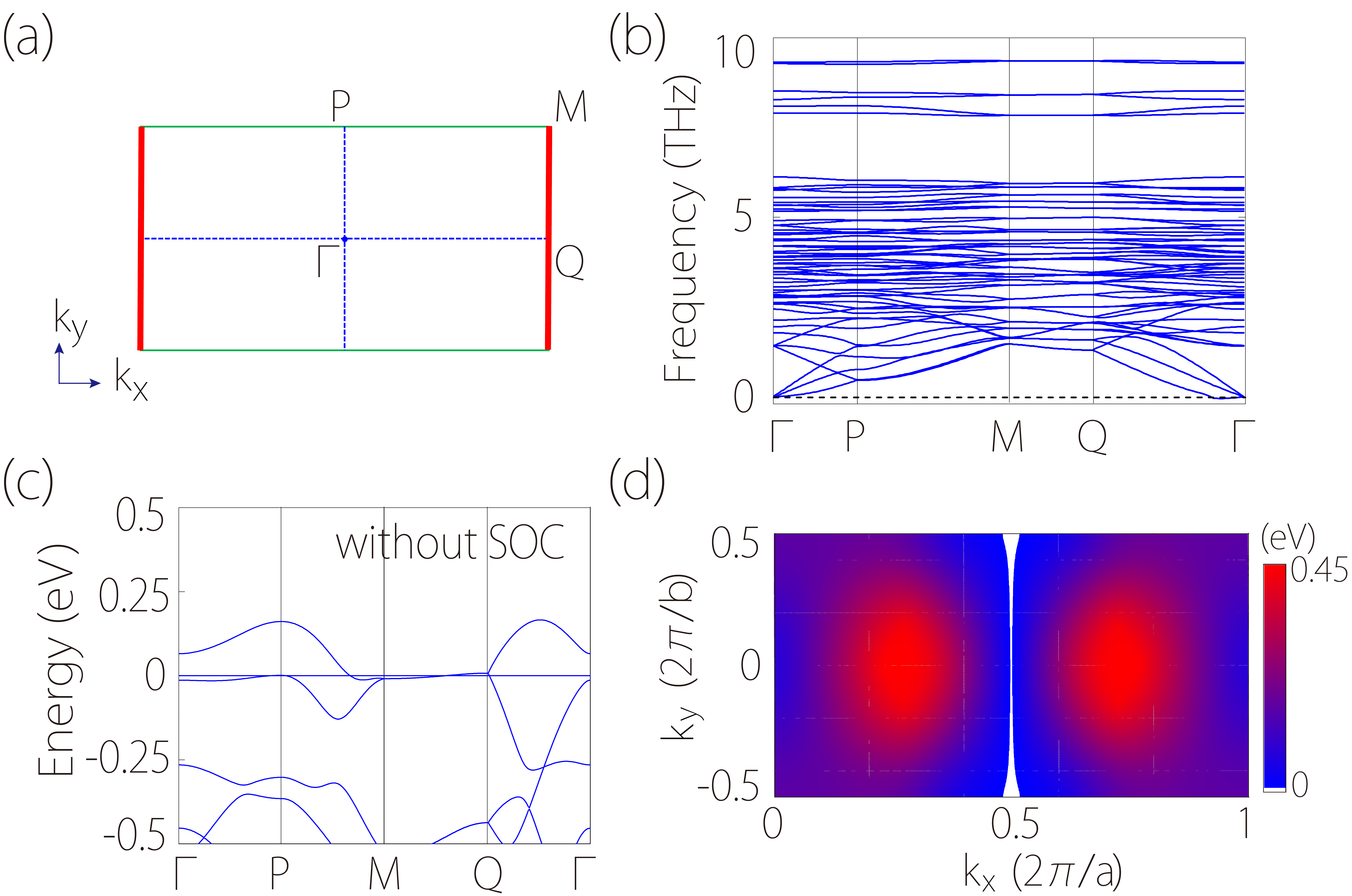}
\caption{(a) Brillouin zone of the monolayer structure. The red color indicates the nodal line in the absence of SOC. (b) Phonon spectrum of the monolayer Ta$_3$SiTe$_6$. (c) Band structure of monolayer Ta$_3$SiTe$_6$ in the absence of SOC. (d) Color-map showing the local gap between the two low-energy bands near the Fermi level. The nodal line along the M-Q path can be observed. }
\label{fig6}
\end{figure}

The band structure of monolayer Ta$_3$SiTe$_6$ in the absence of SOC is shown in Fig.~\ref{fig6}(c). One observes that the system is metallic, and a degenerate and almost flat band appears along the M-Q path at the BZ boundary. It is formed by the linear crossing between two bands and hence represents a nodal line for a 2D system [Fig.~\ref{fig6}(d)]. The existence of this nodal line is dictated by the nonsymmorphic space group symmetry. One notes that the path M-Q: $(\pi,k_y,0)$ is an invariant subspace of the anti-unitary symmetry $\widetilde{\mathcal{M}}_{y}\mathcal{T}$. Similar to the analysis for the bulk case, one finds that $(\widetilde{\mathcal{M}}_{y}\mathcal{T})^2=-1$ on M-Q, hence it requires a Kramers-like double degeneracy for any $k$ point on M-Q. Away from this path, the $\widetilde{\mathcal{M}}_{y}\mathcal{T}$ symmetry is lost (the generic $k$ point is not invariant under $\widetilde{\mathcal{M}}_{y}\mathcal{T}$), so the degeneracy would be lifted. This leads to the nodal line on the M-Q path, as found from the DFT calculation.

\subsection{In the presence of SOC: essential Dirac points}

\begin{figure}[t!]
\includegraphics[width=8.6cm]{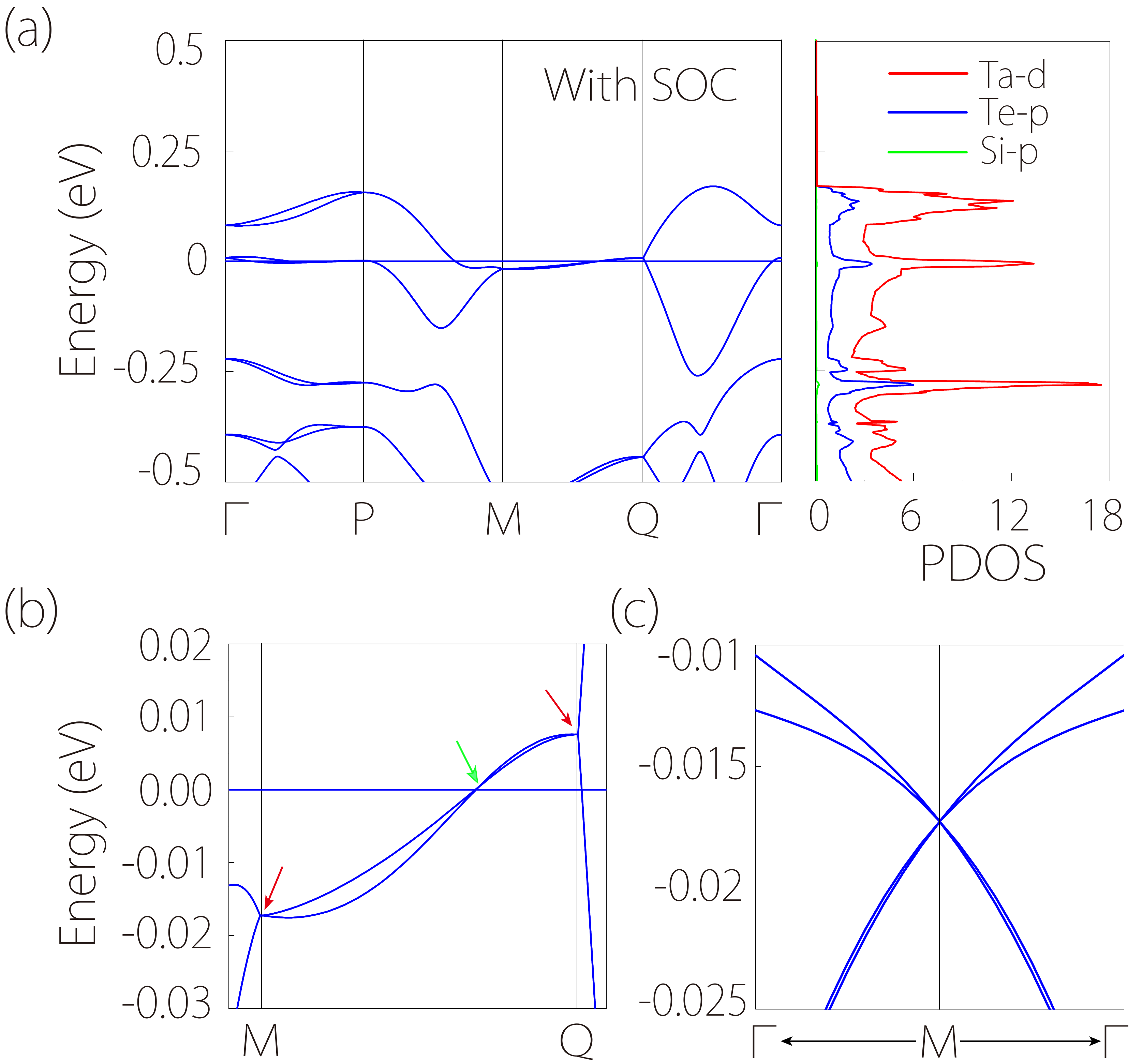}
\caption{(a) Band structure of monolayer Ta$_3$SiTe$_6$ with SOC included. The right panel shows the PDOS. (b) The zoom-in band structure, showing the two Dirac points at M and Q indicated by the red arrows. The green arrow indicates another type-II Dirac point on the M-Q path. (c) Band structure around M along the M-$\Gamma$ direction.}
\label{fig7}
\end{figure}

When the SOC is included, the bands would generally split for a generic $k$ point due to the broken $\mathcal{P}$ symmetry for the monolayer. However, there also remain nontrivial degeneracies along certain high-symmetry paths. The band structure with SOC included is shown in Fig.~\ref{fig7}. Once can observe that: (i) The bands along the M-Q paths (corresponding to the original nodal line in the absence of SOC) have a small splitting; (ii) each band along the P-M, $\Gamma$-Q, and M-Q paths still has a twofold degeneracy; (iii) 2D Dirac points appear at M and Q points, which are fourfold degenerate and have linear dispersions.

The degeneracies in features (ii) and (iii) are resulted from the nonsymmorphic symmetries, as we explain below. First, the double degeneracy on M-Q can be argued in a similar way as before. This path is invariant under the anti-unitary symmetry $\widetilde{\mathcal{M}}_{y}\mathcal{T}$. It is important to note that $(\widetilde{\mathcal{M}}_{y}\mathcal{T})^2=-1$ still hods in the presence of SOC, because although $\mathcal{T}^2=-1$ in this case, $(\widetilde{\mathcal{M}}_{y})^2$ also gets a reversed sign from the operation on spin. This guarantees a Kramers-like double degeneracy on M-Q.

Next, for the double degeneracy on P-M and $\Gamma$-Q, first note that every $k$ point in the BZ is invariant under $\mathcal{M}_{z}$ operation, so each Bloch state can be chosen as an eigenstate of $\mathcal{M}_{z}$. Since
\begin{equation}
({\mathcal{M}}_{z})^2=\overline{E}=-1,
\end{equation}
the $\mathcal{M}_{z}$ eigenvalues must be $g_z=\pm i$. Consider the paths P-M and $\Gamma$-Q. They are also invariant under the symmetry operation $\widetilde{\mathcal{M}}_{y}$. The general commutation relation between $\widetilde{\mathcal{M}}_{y}$ and $\mathcal{M}_{z}$ reads
\begin{equation}\label{eq9}
\mathcal{M}_{z}\widetilde{\mathcal{M}}_{y}=-\widetilde{\mathcal{M}}_{y}\mathcal{M}_{z},
\end{equation}
where the minus sign comes from the anti-commutativity between the operations on spin. Consequently, an $\mathcal{M}_{z}$ eigenstate $|u\rangle$ at a $k$ point on P-M or $\Gamma$-Q must have another degenerate partner $\widetilde{\mathcal{M}}_{y}|u\rangle$ with the opposite $g_z$ eigenvalue. This proves the double degeneracy on P-M and $\Gamma$-Q.

Now we turn to the fourfold degeneracy at M and Q. These high-symmetry points are invariant under all the symmetries ${\mathcal{M}}_{z}$, $\widetilde{\mathcal{M}}_{y}$, and $\mathcal{T}$. Importantly, one notes that for a state $|u\rangle$ at M or Q with an $\mathcal{M}_{z}$ eigenvalue $g_z$, its Kramers partner $\mathcal{T}|u\rangle$ has eigenvalue $-g_z$ (because $g_z=\pm i$). Meanwhile, we have
\begin{equation}\label{eq9}
\mathcal{M}_{z}(\widetilde{\mathcal{M}}_{y}\mathcal{T}|u\rangle)=g_z(\widetilde{\mathcal{M}}_{y}\mathcal{T}|u\rangle),
\end{equation}
which shows that the two orthogonal states $|u\rangle$ and $\widetilde{\mathcal{M}}_{y}\mathcal{T}|u\rangle$ have the same $g_z$.
Therefore, the following four states $\{|u \rangle, \mathcal{T}|u\rangle, \widetilde{\mathcal{M}}_{y}|u \rangle, \widetilde{\mathcal{M}}_{y}\mathcal{T}|u\rangle\}$ are linearly independent and degenerate with the same energy. All the states at M and Q are thus grouped into quartets.

\begin{figure}[tb]
\includegraphics[width=8.6cm]{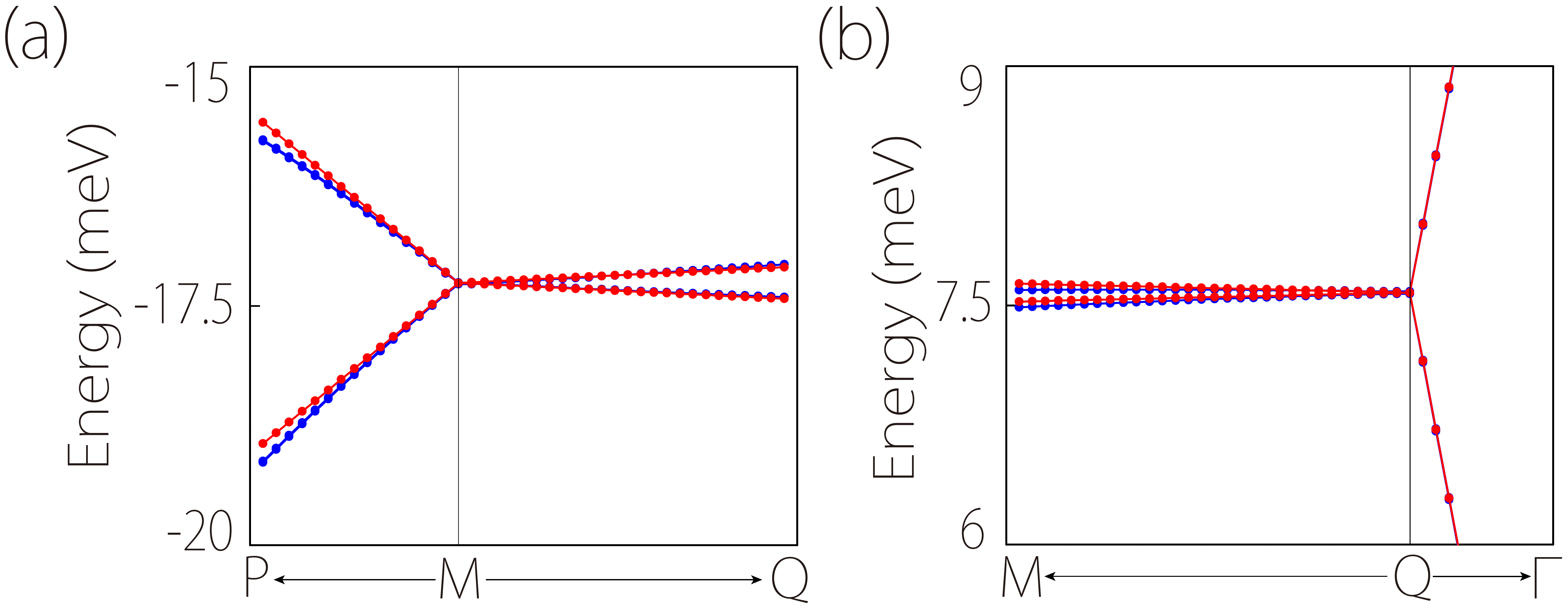}
\caption{Dispersions around the Dirac point fitted by the effective model. (a) is for the M point, and (b) is for the Q point.  The blue dots are results from DFT calculations, and the red dots are the fitting by the $k\cdot p$ model in Eq.~(\ref{kp}).}
\label{fig8}
\end{figure}

To further characterize Dirac points, we construct the $ k\cdot  p$ effective model around each point. The form of the model is determined by the symmetries at M and Q, which include $\mathcal{M}_{z}$, $\widetilde{\mathcal{M}}_{y}$, and $\mathcal{T}$.
Their matrix representations can be obtained from standard references~\cite{Bradley1972}, with $\mathcal{M}_{z}=i\sigma_{0}\tau_{z}$, $\widetilde{\mathcal{M}}_{y}=\sigma_{0}\tau_{x}$, and $\mathcal{T}=-i\sigma_{y}\tau_{0}\mathcal K$, where $\mathcal K$ is the complex conjugation operator, $\sigma_i$ and $\tau_i$ are the Pauli matrices ($\sigma_{0}$ and $\tau_{0}$ are the $2\times 2$ identity matrices) acting on the eigenspace span by the quartet basis. We find that the effective models around M and Q up to linear order share the same form given by
\begin{equation}\label{kp}
\begin{split}
\mathcal{H}=v_xk_x(\sin\theta_x\cos\phi_x\sigma_x+\sin\theta_x\sin\phi_x\sigma_y+\cos\theta_x\sigma_z)\tau_0
  \\
  +v_yk_y(\sin\theta_y\cos\phi_y\sigma_x+\sin\theta_y\sin\phi_y\sigma_y+\cos\theta_y\sigma_z)\tau_z.
\end{split}
\end{equation}
Here, the wave-vector and the energy are measured from the respective Dirac point, and $v_i, \theta_i,\phi_i$, ($i=x,y$) are the model parameters. This low-energy effective model is expanded at M (or Q) and is solely determined by symmetry, it fully describes the type of dispersion around the M (or Q) point, not limited to a particular path.
The model can be used to fit the DFT band structure. Figure~\ref{fig8} shows the fitting result, indicating a very good agreement between the model and the DFT result. The obtained Fermi velocities are $v_x=6.93\times10^3$ m/s and $v_y=7.39\times10^2$ m/s for M; and $v_x=3.99\times10^4$ m/s and $v_y=3.13\times10^2$ m/s for Q.

Unlike the Dirac points in most 2D materials studied so far (here ``2D" refers to the dimension of the system), the nonsymmorphic Dirac points here are intrinsically robust against SOC~\cite{Young2015a,Guan2017}. In fact, they only appear when SOC is present. Such type of Dirac points were initially proposed by Young and Kane~\cite{Young2015a}, and were recently predicted to exist in a realistic material system---the monolayer HfGeTe family materials~\cite{Guan2017}. Here, we provide another candidate. Moreover, the Dirac points here have an important difference from those in monolayer HfGeTe: Here, the inversion symmetry is broken, such that the band degeneracy is completely lifted around the Dirac point (except on the high-symmetry paths). This can be clearly observed in Fig.~\ref{fig7}(c).

In addition, we also note that there is a type-II Dirac point on the M-Q path, as indicated by the green arrow in Fig.~\ref{fig7}(b). Around this type-II Dirac point, the spectrum is completely tipped over along the $k_y$-direction~\cite{Soluyanov2015,Xu2015}. This crossing is protected by the $\mathcal{M}_z$ symmetry, because the two crossing bands (each is doubly degenerate due to the $\widetilde{\mathcal{M}}_{y}\mathcal{T}$ symmetry) have the opposite $\mathcal{M}_z$ eigenvalues.

\section{Discussion and Conclusion}

\begin{figure}[t!]
\includegraphics[width=8.6cm]{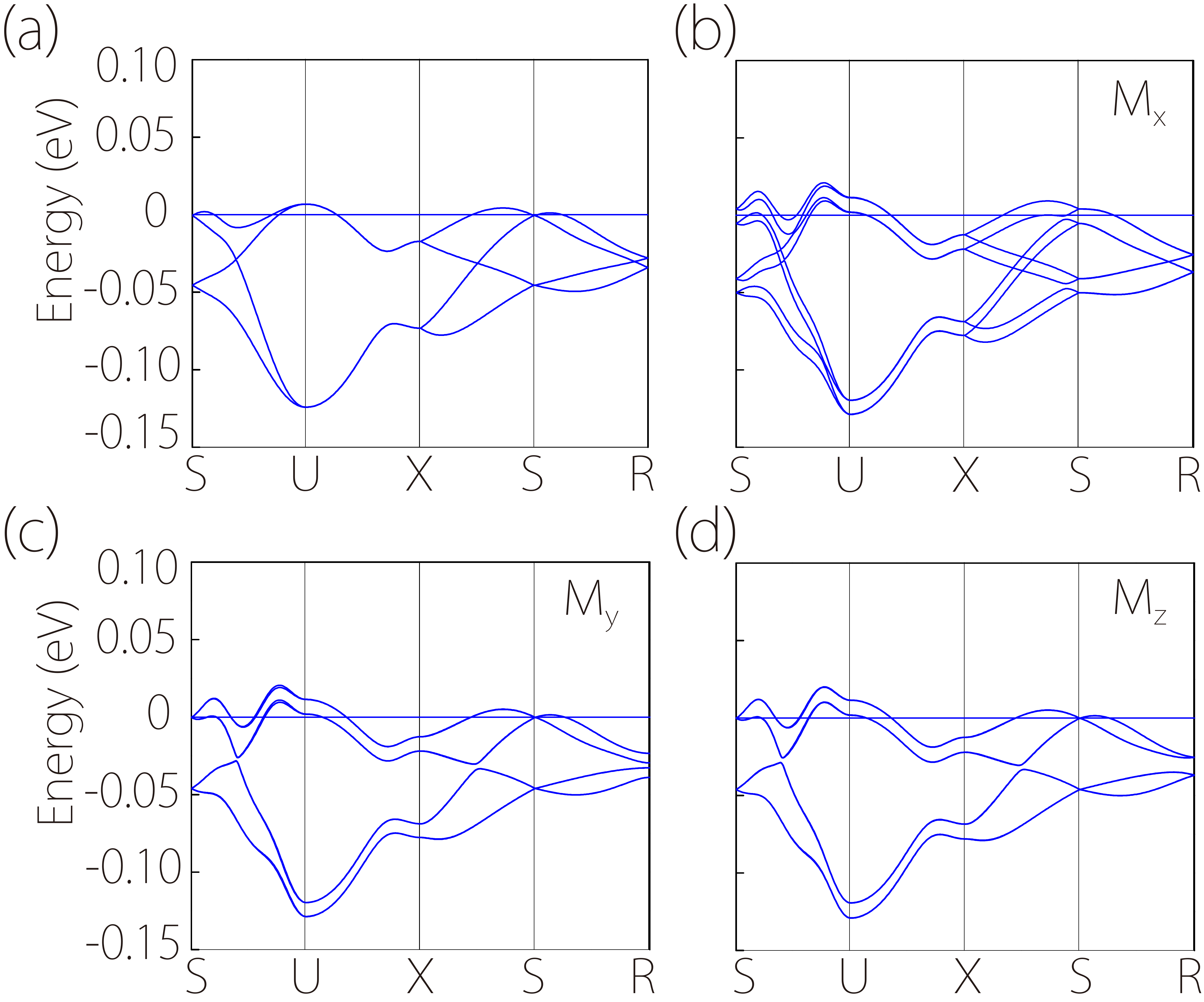}
\caption{(a) Band structure for 3D Ta$_3$SiTe$_6$ without a Zeeman field. (b-d) Band structures with a Zeeman field along the (b) $x$, (c) $y$, or (d) $z$ direction. The value of the Zeeman coupling energy is taken to be 0.005 eV. }
\label{fig9}
\end{figure}

In this work, we have analyzed several types of nontrivial band-crossings that appear in the materials Ta$_3$SiTe$_6$ and Nb$_3$SiTe$_6$. Except for the nodal loop that appears in the bulk structure without SOC, all other band-crossings are of essential type, i.e., their existences (including their locations in BZ) are entirely determined by system symmetry, of which the nonsymmorphic symmetries play a crucial role. Consequently, these band-crossings are quite robust since they do not rely on the band inversion mechanism. In addition, the analysis here can be directly applied for systems with similar symmetries, especially for those materials with space group No.~62 and 26.

On the other hand, it should be noted that although these band-crossings are essential, their energies are not guaranteed to be close to the Fermi level. As we have emphasized in the Introduction Section, it is crucial for such nontrivial band-crossings to be close to the Fermi level for the manifestation of the interesting physics. The materials Ta$_3$SiTe$_6$ and Nb$_3$SiTe$_6$ satisfy this requirement. All the discussed band-crossings are close to the Fermi level (less than 0.1 eV). And the nodal lines and nodal loops have very small energy variations in BZ. Another obvious advantage of the two materials is that they have already been realized in experiment, and ultrathin layers of Nb$_3$SiTe$_6$ have also been demonstrated by micro-exfoliation methods from bulk samples~\cite{hu2015}, which will greatly facilitate the experimental studies on them.

The hourglass Dirac loop in the 3D bulk is quite interesting. The hourglass-type dispersion was previously discussed for 2D systems, as surface states of 3D topological insulators KHg$X$ ($X=$As, Sb, Bi)~\cite{Wang2016a,Ma2017}. The possibility of hourglass dispersion in the 3D bulk systems was also discussed in perovskite iridate materials~\cite{Chen2015a} and in model studies~\cite{Ezawa2016,Wang2017a}. Interestingly, with multiple nonsymmorphic symmetries, it is possible to realize hourglass Weyl or Dirac chains in the reciprocal space~\cite{Bzdusek2016,Wang2017}. In previous examples which are mostly oxide materials, the low-energy band structures are typically complicated, with many bands crossing the Fermi level. In comparison, the band structures of Ta$_3$SiTe$_6$ and Nb$_3$SiTe$_6$ studied here are relatively simple. The hourglass spectrum is close to the Fermi level and should be readily resolved in experiment.

\begin{figure*}[tb]
\includegraphics[width=14cm]{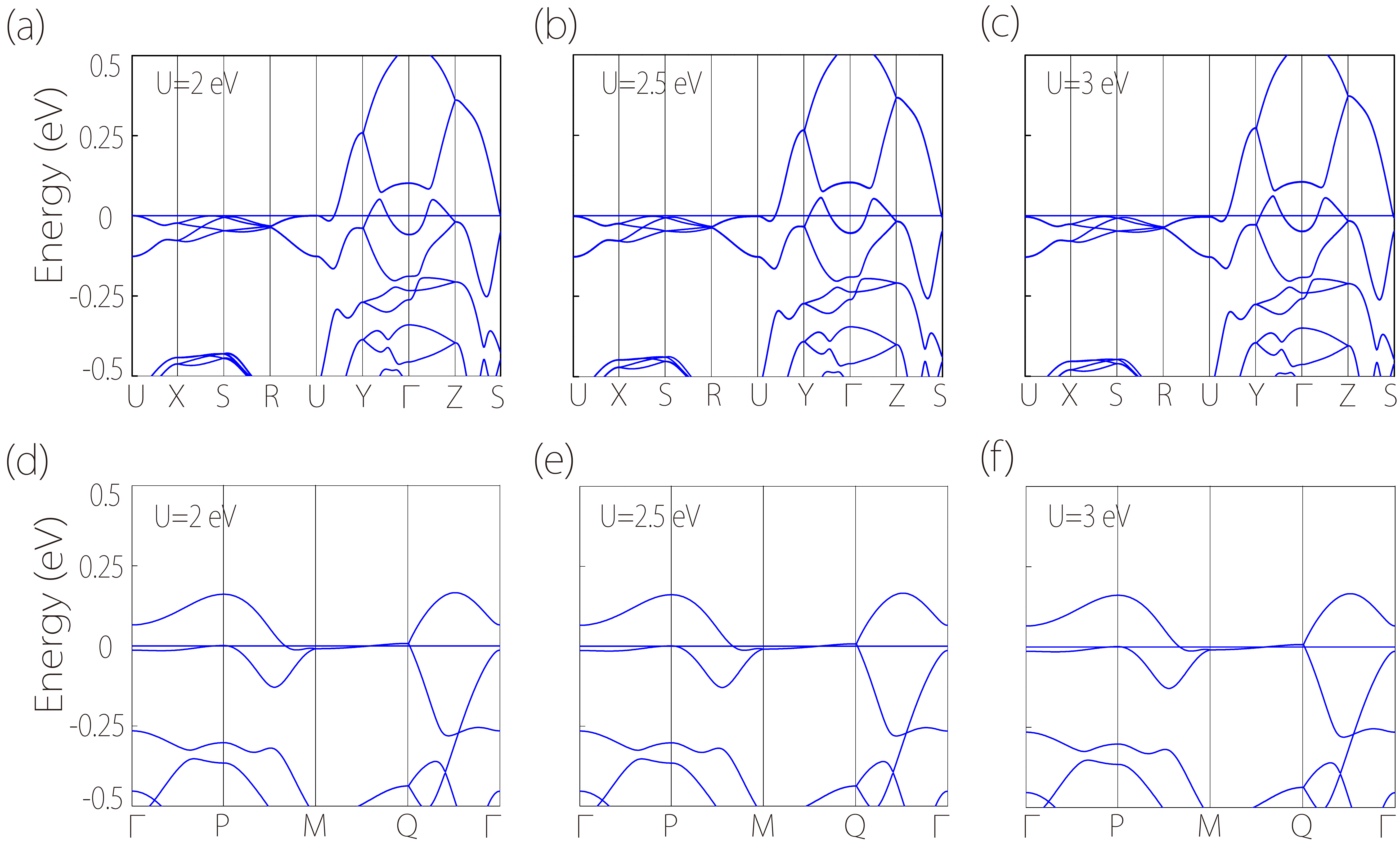}
\caption{Effects of the Hubbard $U$ correction. (a), (b), and (c) are the GGA$+U$ band structures of the 3D Ta$_3$SiTe$_6$ with three different $U$ values (SOC included). (d), (e), and (f) are the corresponding results for the monolayer Ta$_3$SiTe$_6$ (without SOC).}
\label{fig10}
\end{figure*}

We mention that the hourglass Dirac loop could be an interesting playground to study topological phase transitions via symmetry breaking. For example, we consider the effect of Zeeman coupling on the hourglass Dirac loop in Ta$_3$SiTe$_6$. The Zeeman coupling may be realized by external magnetic field or by magnetic doping. As shown in Fig.~\ref{fig9}(b), we find that when the Zeeman field is in the $x$-direction (which is perpendicular to the nodal loop plane), the single hourglass along a path from S to a point on X-U or U-R paths will split into two copies, forming four Weyl loops in the $k_x=\pi$ plane. One also notes that on S-R ($k_x=\pi, k_y, k_z=\pi$) the hourglass does not split, because $\widetilde{\mathcal{M}}_{y}\mathcal{T}$ symmetry is still preserved which dictates the Krammers-like degeneracy on this path.  However, when the Zeeman field is in the $y$-direction or the $z$-direction (which is parallel to the loop plane), the hourglass Dirac loop will be gapped and disappear [see Figs.~\ref{fig9}(c) and \ref{fig9}(d)].

In experiment, the discussed band-crossing features (including the surface states) can be directly probed via the angle-resolved photoemission spectroscopy (ARPES). For the 3D bulk, it has been proposed that under a magnetic field in the plane of the hourglass Dirac loop, there will appear a flat Landau band at the loop energy~\cite{Rhim2015}. It will generate a huge DOS peak that can be probed by the tunneling transport experiment or by scanning tunneling spectroscopy (STS). As for the monolayer, since the SOC splitting is very small, they can be essentially treated as 2D nodal line materials. The nodal line leads to a peak in DOS [see Fig.~\ref{fig7}(a)] close to Fermi energy, which can be detected by STS experiment. In addition, for a nodal line running along the $k_y$ direction (M-Q), one expects strong anisotropy in transport properties: the mobility along the $x$-direction should be much higher due to the Dirac dispersion along this direction.

In conclusion, based on symmetry analysis and first-principles calculations, we have revealed interesting band-crossing features in layered ternary telluride compounds Ta$_3$SiTe$_6$ and Nb$_3$SiTe$_6$. We find that in the absence of SOC, these bulk materials host accidental Dirac loops and essential fourfold nodal lines. In the presence of SOC, there appears an hourglass Dirac loop in the boundary plane of the BZ. The loop has a fourfold degeneracy and each point on the loop is a neck-point of an hourglass dispersion along a certain path. Nontrivial band-crossings also appear when the materials are thinned down to monolayers, including the essential 2D nodal lines in the absence of SOC and the essential 2D Dirac points in the presence of SOC. Our results provide a realistic platform to investigate a variety of topological metal phases, especially those essential band-crossings enabled by nonsymmorphic symmetries. With the materials Ta$_3$SiTe$_6$ and Nb$_3$SiTe$_6$ existing and their ultrathin layers demonstrated in experiment, we expect that the predictions here can be readily verified in experiment in the near future.

\begin{acknowledgements}
The authors thank Weikang Wu, X. C. Wu, and D.L. Deng for valuable discussions. {This work was supported by the National Key R$\&$D Program of China (Grant No. 2016YFA0300600), the MOST Project of China (Grant No. 2014CB920903), the NSF of China
(Grant No. 11734003 and 11574029)}, and the Singapore Ministry of Education AcRF Tier 1 (SUTD-T1-2015004) and Tier 2 (MOE2015-T2-2-144).
\end{acknowledgements}

\begin{appendix}

\begin{figure}[t]
\includegraphics[width=8cm]{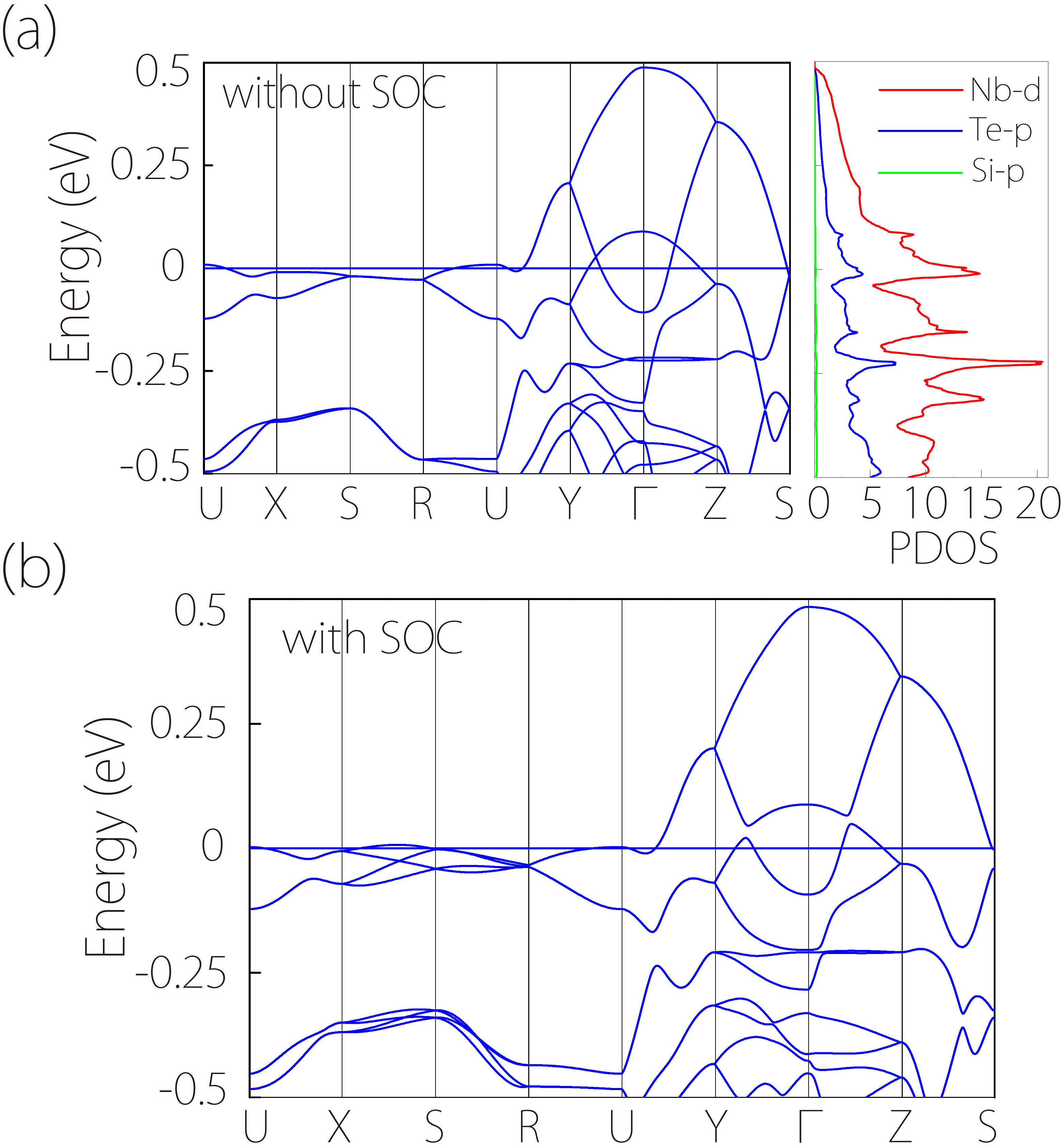}
\caption{Band structures of the 3D Nb$_3$SiTe$_6$ (a) in the absence of SOC, and (b) in the presence of SOC. }
\label{fig11}
\end{figure}

\begin{figure}[t]
\includegraphics[width=8cm]{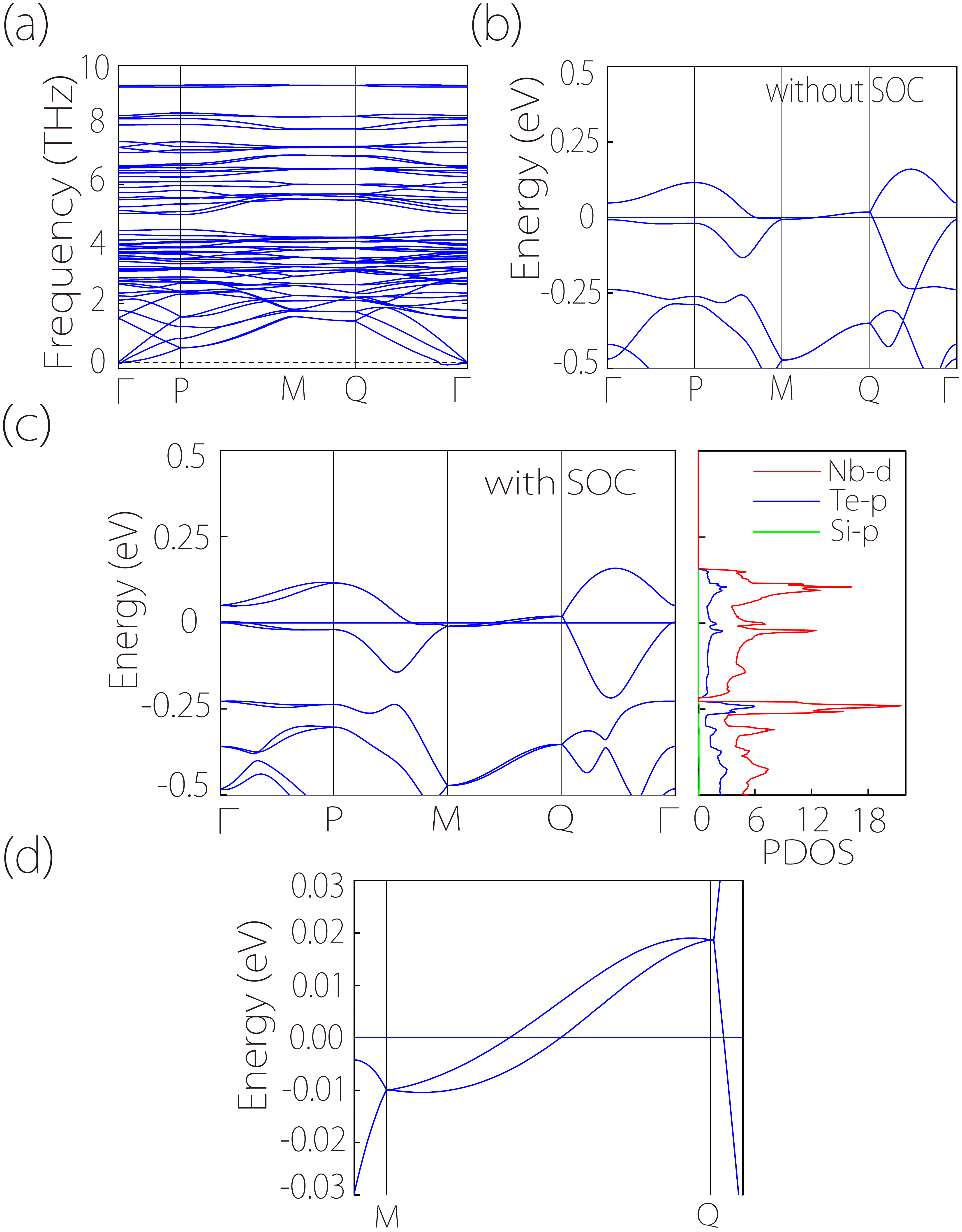}
\caption{(a) Phonon spectrum of the monolayer Nb$_3$SiTe$_6$. Band structures of the monolayer Nb$_3$SiTe$_6$ (b) in the absence of SOC, and (c) in the presence of SOC. The right panel in (c) shows the PDOS. (d) The zoom-in image of the band structure in (c) along the path M-Q. }
\label{fig12}
\end{figure}

\section{Fourfold degeneracy on R-U and Z-S in the presence of SOC}

In Sec.~\ref{3DwSOC}, we have analyzed the fourfold degeneracy on U-X for the 3D bulk band structure in the presence of SOC. Here we present the detailed analysis of the fourfold degeneracy on the other two paths R-U and Z-S.

First consider the path R-U: $(\pi,\pi,k_z)$, where $-\pi < k_z \le \pi$. It is an invariant subspace of $\widetilde{\mathcal{M}}_{x}$. so each Bloch state $|u\rangle$ there can be chosen as an eigenstate of $\widetilde{\mathcal{M}}_{x}$. Since
\begin{equation}
(\widetilde{\mathcal{M}}_{x})^2=T_{011}\overline{E},
\end{equation}
which is equal to $e^{-ik_z}$ on R-U, the $\widetilde{\mathcal{M}}_{x}$ eigenvalues are given by $g_x=\pm e^{-ik_z/2}$. From the commutation relation
\begin{equation}
\widetilde{\mathcal{M}}_{x}\mathcal{P}=T_{111}\mathcal{P}\widetilde{\mathcal{M}}_{x},
\end{equation}
one finds that along R-U,
\begin{equation}\label{eq3}
\widetilde{\mathcal{M}}_{x}(\mathcal{P}\mathcal{T}|u\rangle)=g_x(\mathcal{P}\mathcal{T}|u\rangle),
\end{equation}
indicating that the state $|u\rangle$ and $\mathcal{P}\mathcal{T}|u\rangle$ have the same eigenvalue $g_x$.

In addition, R-U is also invariant under $\widetilde{\mathcal{M}}_{y}$. One finds that $\{\widetilde{\mathcal{M}}_{x},\widetilde{\mathcal{M}}_{y}\}=0$ along R-U,
so the two states $|u\rangle$ and $\widetilde{\mathcal{M}}_{y}|u\rangle$ have opposite $\widetilde{\mathcal{M}}_{x}$ eigenvalues. Therefore, at each $k$ point on R-U, the four
states $\{|u\rangle,\mathcal{P}\mathcal{T}|u\rangle,\widetilde{\mathcal{M}}_{y}|u\rangle,
\widetilde{\mathcal{M}}_{y}\mathcal{P}\mathcal{T}|u\rangle\}$
must be linearly independent and degenerate with the same energy.

Next, for the path Z-S: $(k_x,0,\pi)$, where $-\pi < k_x \le \pi$, we can choose the Bloch states to be eigenstates of  $\widetilde{\mathcal{M}}_{y}$, with eigenvalues $g_y=\pm ie^{-ik_x/2}$. From the relation
\begin{equation}
\widetilde{\mathcal{M}}_{y}\mathcal{P}=T_{110}\mathcal{P}\widetilde{\mathcal{M}}_{y},
\end{equation}
we have on Z-S
\begin{equation}
\widetilde{\mathcal{M}}_{y}(\mathcal{P}\mathcal{T}|u\rangle)=-g_y(\mathcal{P}\mathcal{T}|u\rangle).
\end{equation}
This shows that the states $|u\rangle$ and $\mathcal{P}\mathcal{T}|u\rangle$ have the opposite $g_y$ eigenvalues. In addition, Z-S is invariant under the anti-unitary symmetry $\widetilde{\mathcal{M}}_{x}\mathcal{T}$, which generates a Kramers-like degeneracy since $(\widetilde{\mathcal{M}}_{x}\mathcal{T})^2=-1$. Note that
\begin{equation}\label{eq4}
\widetilde{\mathcal{M}}_{y}\widetilde{\mathcal{M}}_{x}=-T_{1\bar10}
\widetilde{\mathcal{M}}_{x}\widetilde{\mathcal{M}}_{y},
\end{equation}
such that on Z-S, we have
\begin{equation}\label{eq6}
\widetilde{\mathcal{M}}_{y}(\widetilde{\mathcal{M}}_{x}\mathcal{T}|u\rangle)=g_y
(\widetilde{\mathcal{M}}_{x}\mathcal{T}|u\rangle).
\end{equation}
Hence, the degenerate pair $|u\rangle$ and $\widetilde{\mathcal{M}}_{x}\mathcal{T}|u\rangle$ have same $\widetilde{\mathcal{M}}_{y}$ eigenvalue. Therefore, the four linearly independent states $\{|u\rangle,\mathcal{P}\mathcal{T}|u\rangle,\widetilde{\mathcal{M}}_{x}\mathcal{T}|u\rangle,
\mathcal{P}\widetilde{\mathcal{M}}_{x}|u\rangle\}$
form a degenerate quartet on Z-S.

\section{Band structure with Hubbard U correction}

In order to test the effect of electron correlation on the band structure, we performed the GGA$+U$ calculations, taking into account the Hubbard U corrections on the transition metal elements. Figure~\ref{fig10} shows the obtained results for Ta$_3$SiTe$_6$.
Here, we have tested the $U$ values for the Ta $d$-orbitals up to 3 eV. The results show that there is little change compared from the GGA results for both the 3D bulk and the monolayer structures.

\section{Results for Nb$_3$SiTe$_6$}

Nb$_3$SiTe$_6$ share the same type of lattice structure as Ta$_3$SiTe$_6$. For the 3D bulk structure, the optimized lattice parameters are $a=6.378$ \AA, $b=11.553$ \AA, and $c=13.994$ \AA, which is in good agreement with the experimental values ($a=6.353$ \AA, $b=11.507$ \AA, and $c=13.938$ \AA)~\cite{li1992}. For the monolayer structure, the optimized lattice parameters are $a=6.410$ \AA, $b=11.611$ \AA. The monolayer structure is also found to be dynamically stable from the calculated phonon spectrum [see Fig.~\ref{fig12}(a)]. The band structure results for the 3D bulk and the monolayer structures are shown in Fig.~\ref{fig11} and Fig.~\ref{fig12}, respectively. One observes that they share the similar features as those discussed in the main text for Ta$_3$SiTe$_6$.

\end{appendix}


%


\end{document}